\documentstyle[12pt,epsf]{article} 

\newcommand{\be}{\begin{equation}} 
\newcommand{\en}{\end{equation}}

\begin{document} 

\setlength{\parindent}{0cm}
\setlength{\parskip}{17pt}
\addtolength{\textheight}{0.5cm}
  
\centerline{\large\bf Deconfinement in SU(2) Yang-Mills theory as a}
\centerline{\large\bf center vortex percolation transition}
  
\bigskip
\centerline{M.~Engelhardt, K.~Langfeld, H.~Reinhardt and O.~Tennert}
\vspace{0.2 true cm} 
\centerline{\em Institut f\"ur Theoretische Physik, Universit\"at 
T\"ubingen }
\centerline{\em D--72076 T\"ubingen, Germany}

\bigskip
  
\begin{abstract}
By fixing lattice Yang-Mills configurations to the maximal center gauge
and subsequently applying the technique of center projection, one can
identify center vortices in these configurations. Recently, center
vortices have been shown to determine the string tension between
static quarks at finite temperatures (center dominance); also, they
correctly reproduce the deconfining transition to a phase with
vanishing string tension. After verifying center dominance also for
the so-called spatial string tension, the present analysis focuses
on the global topology of vortex networks. General arguments are given
supporting the notion that the deconfinement transition in the center
vortex picture takes the guise of a percolation transition. This 
transition is detected in Monte Carlo experiments by concentrating on various
slices through the closed vortex surfaces; these slices, representing
loops in lattice universes reduced by one dimension, clearly exhibit
the expected transition from a percolating to a non-percolating, deconfined,
phase. The latter phase contains a large proportion of vortex loops
winding around the lattice in the Euclidean time direction. At the
same time, an intuitive picture clarifying the persistence of the spatial
string tension in the deconfined phase emerges.
\end{abstract}



\newpage

\addtolength{\baselineskip}{2pt}

\section{Introduction}
The description of hadronic matter in terms of confined quark and
gluon constituents carrying a color quantum number has opened the
prospect of a new, deconfined, phase of matter in which colored
excitations can propagate over distances much larger than typical
hadronic sizes. In the framework of pure Yang-Mills theory, the
transition to this new phase is thought to occur as a function of
temperature. While compelling evidence for the deconfining phase 
transition has been collected in lattice Monte Carlo simulations 
\cite{mm},\cite{rothe}, it is necessary to concomitantly develop an 
intuitive picture for the deconfinement phenomenon in order to be able 
to treat scenaria as complex as heavy ion collisions; such collision 
experiments, planned at RHIC and LHC, are hoped to produce lumps of 
deconfined matter in the near future.

The question of the deconfinement transition can not be separated
from an underlying picture of the confinement mechanism itself.
Conversely, any purported mechanism of confinement should also be
able to incorporate deconfinement. The present paper concentrates on the
center vortex picture of confinement in the case of SU(2) color. This
mechanism, initially proposed in \cite{thoo}-\cite{corn79},
generates an area law for the Wilson loop by invoking the presence of
vortices in typical configurations entering the Yang-Mills functional
integral. These vortices are closed two-dimensional surfaces in
four-dimensional space-time, or, equivalently, closed lines in the
three dimensions making up, e.g., a time slice. They carry flux such
that they contribute a factor corresponding to a nontrivial center 
element of the gauge group to any Wilson loop whenever they pierce its
minimal area; in the case of SU(2) color to be treated below, that is a
factor $-1$. If the vortices are distributed in space-time sufficiently
randomly, then samples of the Wilson loop of value $+1$ (originating
from loop areas pierced an even number of times by vortices) will
strongly cancel against samples of the Wilson loop of value $-1$
(originating from loop areas pierced an odd number of times by
vortices), generating an area law fall-off. The simplest (SU(2)) model
visualization which demonstrates this is the following: Consider
a universe of volume $L^4 $, and a two-dimensional slice through it
of area $L^2 $, containing a Wilson loop spanning an area $A$. Generical 
vortices will pierce the slice at points; assume $N$ of these points to 
be randomly distributed on the slice. Then the probability of finding 
$n$ such points inside the Wilson loop area is binomial,
\begin{equation}
P_N (n) = \left( \begin{array}{c} N \\ n \end{array} \right) 
\left( \frac{A}{L^2 } \right)^{n}
\left( 1 - \frac{A}{L^2 } \right)^{N-n}
\end{equation}
and the expectation value of the Wilson loop becomes
\begin{equation}
\langle W \rangle = \sum_{n=0}^{N} (-1)^{n} P_N (n)
=\left( 1 - \frac{2\rho A}{N} \right)^{N}
\stackrel{N\rightarrow \infty }{\longrightarrow } e^{-2\rho A}
\end{equation}
where the planar density of the intersection points $\rho =N/L^2 $
is kept constant as $N\rightarrow \infty $.
One thus obtains an area law with string tension $\kappa = 2\rho $.
In a more realistic calculation, one would e.g. take into account
interactions between the vortices \cite{corr}; the proportionality
constant $\kappa /\rho $ turns out to be close to $1.4$ in zero 
temperature lattice measurements \cite{giedt},\cite{tempv} (a survey of 
existing data follows further below).

The emphasis of the present work, however, lies not on relatively
short-range properties of the vortices such as their thickness,
but on their long-range topology. This is where the argument
presented above has more serious shortcomings. For one, it suggests
that the expectation value of a Wilson loop might depend on the
area with which one chooses to span the loop. However, due to the closed
nature of the vortices, the choice of area is in fact immaterial,
as it should be. In a more precise, area-independent, manner of 
speaking than adopted above, the value a Wilson loop takes in a
given vortex configuration should be derived from the linking numbers
of the vortices with the loop. Now, the above model visualization
demonstrating an area law implicitly makes a strong assumption about 
the long-range topology of vortex configurations: For the
intersection points of vortices with a given plane to be distributed
sufficiently randomly on the plane to generate confinement, typical 
vortices or vortex networks (note that vortices are not forbidden to 
self-intersect) must extend over the entire universe. Consider the 
converse, namely that there is an upper bound to the space-time 
extension of single vortices or vortex networks. Then an
intersection point of a vortex with a plane always comes paired with
another such point a finite distance away, due to the closed character
of the vortices. This pairing in particular would preclude an area
law for the Wilson loop, as can be seen more clearly with the help
of another simple model.

Consider a universe as above, but with the additional information that 
intersection points of vortices with a two-dimensional slice come in 
pairs at most a distance $d$ apart. Then the only pairs which can
contribute a factor $-1$ to a planar Wilson loop are ones whose midpoints
lie in a strip of width $d$ centered on the trajectory of the loop.
Denote by $p$ the probability that a pair which satisfies this condition
actually does contribute a factor $-1$. This probability is an appropriate
average over the distances of the midpoints of the pairs from the
Wilson loop, their angular orientations, the distribution of separations
between the points making up the pairs, and the local geometry of the
Wilson loop up to the scale $d$. The probability $p$, however, does not
depend on the macroscopic extension of the Wilson loop. A pair which
is placed at random on a slice of the universe of area $L^2 $ has
probability $p\cdot A/L^2 $ of contributing a factor $-1$ to a Wilson
loop, where $A$ is the area of the strip of width $d$ centered on the
Wilson loop trajectory. To leading order, $A=Pd$, where $P$ is the
perimeter of the Wilson loop; subleading corrections are induced by
the local loop geometry. Now, placing $N$ pairs on a slice of the
universe of area $L^2 $ at random, the probability that $n$ of them
contribute a factor $-1$ to the Wilson loop is
\begin{equation}
P_{N_{pair} } (n) = \left( \begin{array}{c} N_{pair} \\ n \end{array} 
\right) \left( \frac{pPd}{L^2 } \right)^{n}
\left( 1 - \frac{pPd}{L^2 } \right)^{N_{pair} -n}
\end{equation}
and, consequently, the expectation value of the Wilson loop for large
universes is
\begin{equation}
\langle W \rangle = \sum_{n=0}^{N_{pair} } (-1)^{n} P_{N_{pair} } (n)
\stackrel{N_{pair} \rightarrow \infty }{\longrightarrow } e^{-\rho pPd}
\end{equation}
where $\rho =2N_{pair} /L^2 $ is the planar density of points. One thus 
observes a perimeter law, negating confinement, if the space-time extension
of vortices or vortex networks is bounded. They must thus extend over
the entire universe, i.e. percolate, in order to realize confinement.

Conversely, therefore, a possible mechanism driving the deconfinement 
transition in the vortex picture is that vortices, in a sense to be made
more precise below, cease to be of arbitrary length, i.e. cease to 
percolate, in the deconfined phase \cite{tempv}.
The main result of the present work is that this is indeed the case, 
implying that the deconfinement transition can be characterized as a 
vortex percolation transition.

Before entering into the details, it should be noted that a description
of the deconfinement transition in terms of percolation phenomena
has also been advocated in frameworks based on Yang-Mills degrees of
freedom other than vortices. For one, electric flux is expected to 
percolate in the {\em deconfined} phase, while it does not percolate
in the confined phase. Note that this is the reverse, or dual, of the
magnetic vortex picture. General arguments related to electric flux
percolation were recently advanced in \cite{satz}; also, specific 
electric flux tube models support this picture \cite{patel}.

On the other hand, in the dual superconductor picture of confinement,
it has been observed that the confined phase is characterized by the 
presence of a magnetic monopole loop percolating throughout the (lattice) 
universe, whereas the monopole configurations are considerably more 
fragmented in the deconfined phase and cease to percolate \cite{borin}. 
To the authors' knowledge, however, this is mainly an
empirical observation and there is no clear physical argument connecting
the deconfinement transition and monopole loop percolation. Indeed,
there has been speculation that the two phenomena may be disconnected
\cite{borin}. This should be contrasted with the vortex language, which,
as discussed at length above, has the advantage of providing a clear
physical picture motivating an interrelation between vortex percolation
and confinement.

\section{Tools and survey of existing data}
Before vortex clustering properties can be investigated in detail,
some technical prerequisites have to be met; foremost, one 
must have a manageable definition of vortices, i.e. an algorithm which 
allows to localize and isolate them in Yang-Mills field configurations. 
After the initial proposal of the center vortex confinement mechanism, 
a first hint of the existence of vortex configurations was provided by 
the Copenhagen vacuum \cite{spag} based on the observation that a constant 
chromomagnetic field in Yang-Mills theory is unstable with respect to the 
formation of flux tube domains in three-dimensional space. Later it was 
observed that the chromomagnetic flux associated with these domains indeed 
is quantized according to the center of the gauge group \cite{spagz}. 
However, the theory of these flux tubes quickly 
became too technically involved to allow e.g. the study of global 
properties of the flux tube networks, especially at finite temperatures. 
In parallel, efforts were undertaken to define and isolate vortices on a 
space-time lattice. One definition, proposed by Mack and coworkers 
\cite{mack} and developed further by Tomboulis \cite{tomold} introduces a 
distinction between thin and thick vortices, only the latter remaining 
relevant in the continuum limit. The defining property of these thick 
vortices is the nontrivial center element factor they contribute to a 
large Wilson loop when they pierce its minimal area. This definition has 
the advantage of being gauge invariant; on the other hand, it does not
allow to easily localize vortices in the sense of associating a 
space-time trajectory with them.

A different line of reasoning has only recently been developed in a series
of papers by Del Debbio et al \cite{giedt},\cite{deb96}-\cite{deb97aug}. 
One chooses a gauge which as much as possible concentrates
the information contained in the field configurations on particular
collective degrees of freedom, in the present case, the vortices.
If this concentration of information is successful (more about this
question further below), one obtains a good
approximation of the dynamics by neglecting the residual deviations
away from the chosen collective degrees of freedom, i.e. by projecting
onto them. This type of approach was pioneered by G.~'t~Hooft, who
introduced the class of Abelian gauges and the subsequent Abelian
projection in order to study Abelian monopole degrees of freedom
\cite{tho81}. In complete analogy, one can introduce maximal center
gauges \cite{giedt},\cite{deb96}-\cite{deb97aug}, in which one uses the 
gauge freedom to choose link variables on a space-time lattice as close as 
possible to center elements of the gauge group. Subsequently, one can 
perform center projection, i.e. replace the gauge-fixed link variables with
the center elements nearest to them on the group. 

Given such a lattice of center elements, i.e. in the case of SU(2) color, 
a lattice with links taking the values $\pm 1$, center vortices are
defined as follows: Consider
all plaquettes in the lattice. If the links bordering the plaquette
multiply to $-1$, then a vortex pierces that plaquette. These are
precisely the vortices needed for the center vortex mechanism of
confinement. To see this, one merely needs to apply Stokes' theorem:
Consider a Wilson loop $W$, made up of links $l=\pm 1$, and an area
$A$ it circumscribes, made up of plaquettes $p=\pm 1$ (the value of
a plaquette is given by the product of the bordering links). Then
\begin{equation}
W = \prod_{l \in W} l = \prod_{p \in A} p
\end{equation}
(the same letter was used here to denote both space-time objects and
the associated group elements). In other words, the Wilson loop receives
a factor $-1$ from every vortex piercing the area. Furthermore, the
product of all plaquettes making up a three-dimensional elementary
cube in the lattice is $1$, since this product contains every link
making up the cube twice. This fact, which in physical terms is a 
manifestation of the Bianchi identity, implies that every such cube has an 
even number of vortices piercing its surfaces; consequently, any projection
of the lattice down to three dimensions contains only closed vortex
lines. Since any cut through a two-dimensional vortex surface in four
dimensions is thus a closed line, the original surface is also closed.
Note that if one defines the dual lattice as a lattice with the same
spacing $a$ as the original one, shifted with respect to the latter
by the vector $(a/2,a/2,a/2,a/2)$, then vortices are made up of 
plaquettes on the dual lattice.

In the work presented here, the specific maximal center gauge called
``direct maximal center gauge'', see e.g. \cite{giedt}, was used. This 
gauge is reached by maximizing the quantity
\begin{equation}
\sum_{l} \left| \mbox{tr} \, U_l \right|^{2} \ ,
\end{equation}
where $l$ labels all the links $U_l $ on the lattice. Center projection
then means replacing
\begin{equation}
U_l \rightarrow \mbox{sign tr } U_l \ .
\end{equation}
In practice, the question whether the gauge fixing and projection 
procedure indeed successfully concentrates the relevant physical 
information on the collective degrees of freedom being projected on 
is difficult to settle a priori; most often, this is tested 
a posteriori by empirical means. Success furthermore depends on the 
specific physics, i.e. the observable, under consideration. One
carries out two Monte Carlo experiments, using the full 
Yang-Mills action as a weight in both cases, and samples
the observable in question, such as e.g. the Wilson loop, using 
either the full lattice configurations or the center projected 
ones. If the results agree, one refers to this state of affairs
as ``center dominance'' for that particular observable.
Center dominance for the Wilson loop is interpreted as
evidence that the center gauge concentrates the physical information 
relevant for confinement on the vortex degrees of freedom, and that 
consequently center projection, i.e. projection onto the associated 
vortex configuration, constitutes a good approximation. Center 
dominance has been verified for the long-range part of the static quark
potential at zero temperature (see \cite{deb96}-\cite{deb97aug} for the 
SU(2) theory and \cite{giedt} for the SU(3) theory).

This recent verification of center dominance has sparked renewed interest
in the vortex picture of confinement. In establishing the relevance of
vortex degrees of freedom for confinement, it provides the necessary basis
for any further investigation of vortex properties. An observation 
analogous to center dominance has been made in the framework of 
the gauge-invariant vortex definition advanced by Tomboulis \cite{tom97}. 
There, one samples both the quantities $W$ and sign$(W)$, $W$ denoting the 
Wilson loop; sign$(W)$ is interpreted as containing only the center vortex 
contributions to $W$, whereas all other fluctuations of the gauge fields 
are neglected. One finds that the expectation value of sign$(W)$ alone
already provides the full string tension, i.e. one finds a gauge-invariant
type of center dominance (see \cite{tom97} for the SU(2) theory
and \cite{tom98} for the SU(3) theory). Subsequently, it has been noted
that this type of center dominance without gauge fixing can in fact be
understood in quite simple terms \cite{ogilv}, and that furthermore
the density of center vortices arising on center-projected lattices
without gauge fixing does not exhibit the renormalization group scaling
corresponding to a finite physical density \cite{faunf}.

In parallel, other vortex properties were investigated. There is
evidence in the SU(2) theory that the vortices defined by center gauging 
and center projection indeed localize thick vortices as defined by 
their center element contributions to linked Wilson loops 
\cite{deb97},\cite{deb97aug}. In both the gauge-fixed and unfixed frameworks, 
absence of vortices was shown to imply absence of confinement 
\cite{deb97},\cite{deb97aug},\cite{tomabs}. In zero temperature lattice
calculations using the maximal center gauge, the planar 
density of intersection points of vortices with a given surface was shown
to be a renormalization group invariant, physical quantity in the 
SU(2) theory, cf. \cite{kurt} (note erratum in \cite{tempv}) and also 
\cite{giedt}. This planar density equals approximately 
$3.6/\mbox{fm}^{2} $ if one fixes the scale by positing a string tension of 
$(440 \mbox{MeV})^{2} $. Also the radial distribution function of these 
intersection points on a plane is renormalization group invariant 
\cite{corr}.  Furthermore, if one takes into account the thickness of 
center vortices, they are able to account for the ``Casimir scaling'' 
behavior of higher representation Wilson loops, a feature which
hitherto was considered incompatible with the vortex confinement
mechanism \cite{fa97},\cite{fa98}. Also, the monopoles generated by 
the maximal Abelian gauge have been found to lie on the center vortices 
identified in a subsequent (indirect) maximal center gauge, forming 
monopole-antimonopole chains \cite{deb97aug}. Recently, a modified
$SU(2)$ lattice ensemble was investigated in which all center vortices
had been removed, with the result that chiral symmetry is restored and
all configurations turn out to belong to the topologically trivial sector
\cite{phil}.

The purpose of the present analysis is to confront the center vortex
picture of confinement with the finite temperature transition to a
deconfined phase observed in Yang-Mills lattice experiments. Some
previous work on vortex properties at finite temperatures has already
been carried out, generalizing the zero-temperature results surveyed
above. For one, the authors reported some preliminary work in
\cite{tempv}. There, center dominance for the string tension between
static quarks was verified at finite temperatures, and the transition
to the deconfined phase with a vanishing string tension observed at the
correct temperature in the center-projected theory. A depletion in the
density of vortex intersection points with a plane extending in the
(Euclidean) time and one space direction occurs as one crosses into
the deconfined phase. The vortices are to a certain extent polarized
in the time direction. However, the polarization is not complete; an
area spanned by a Polyakov loop correlator is still pierced by a
finite density of vortices. Thus, more detailed correlations between
these vortex intersection points must induce the deconfinement transition;
this led the authors to first conjecture in \cite{tempv} that the
deconfinement transition in the center vortex picture may be connected
to global properties of vortex networks such as their connectivity.

Very recently, a related investigation into the global topology of the 
two-dimensional vortex surfaces in four space-time dimensions was
reported in \cite{bertl}, including the case of finite temperatures.
This investigation focused on properties such as orientability and
genus of the surfaces, in particular, changes in these characteristics
as one crosses into the deconfined phase. In the present work, the 
global properties of vortex surfaces are considered from a slightly 
different vantage point, namely specifically with a view to testing the 
heuristic arguments given in the introduction, connecting confinement with 
percolation properties. For this purpose, it will be necessary to consider 
in more detail different slices of vortex surfaces; details follow below.

\section{Spatial string tension}
\label{centdo}
Before doing so, a certain gap in the existing literature on center 
vortices at finite temperatures should be addressed. As already mentioned 
above, the basis for the center vortex picture of confinement is the
empirical observation of {\em center dominance} for the Wilson loop.
Without first establishing center dominance for an observable under
investigation, a more detailed discussion of the manner in which vortex
dynamics influence the observable runs the risk of being largely academic. 
Center dominance for the finite-temperature long-range heavy quark 
potential, via the corresponding Polyakov loop correlator, was verified 
in \cite{tempv}, as mentioned above; however, in what follows, also the 
behavior of the so-called spatial string tension, extracted from large 
spatial Wilson loops, will be under scrutiny. To provide the necessary basis 
for this, center dominance for large spatial Wilson loops should first be 
checked. For this purpose, the authors have carried out lattice 
measurements of spatial Creutz ratios, using center-projected 
configurations to evaluate the Wilson loops, for three temperatures.

Before presenting the results, a comment on the physical scales is
in order. Throughout this paper, the zero-temperature string tension
is taken to be $\kappa = (440  \mbox{MeV})^2 $, the lattice spacing
$a(\beta ) $ at inverse coupling $\beta = 2.3$ is determined by
$\kappa a^2 =0.12$, and one-loop scaling is used for the 
$\beta $-dependence of $a$. The deconfinement temperature is identified
as $T_C =300$ MeV, cf. \cite{tempv}. It should be noted that these scales
are fraught with considerable uncertainty, of the order of $10\% $, due
to finite size effects. This was discussed in more detail in \cite{tempv}.

\begin{figure}

\vspace{-4cm}

\centerline{
\epsfysize=14cm
\epsffile{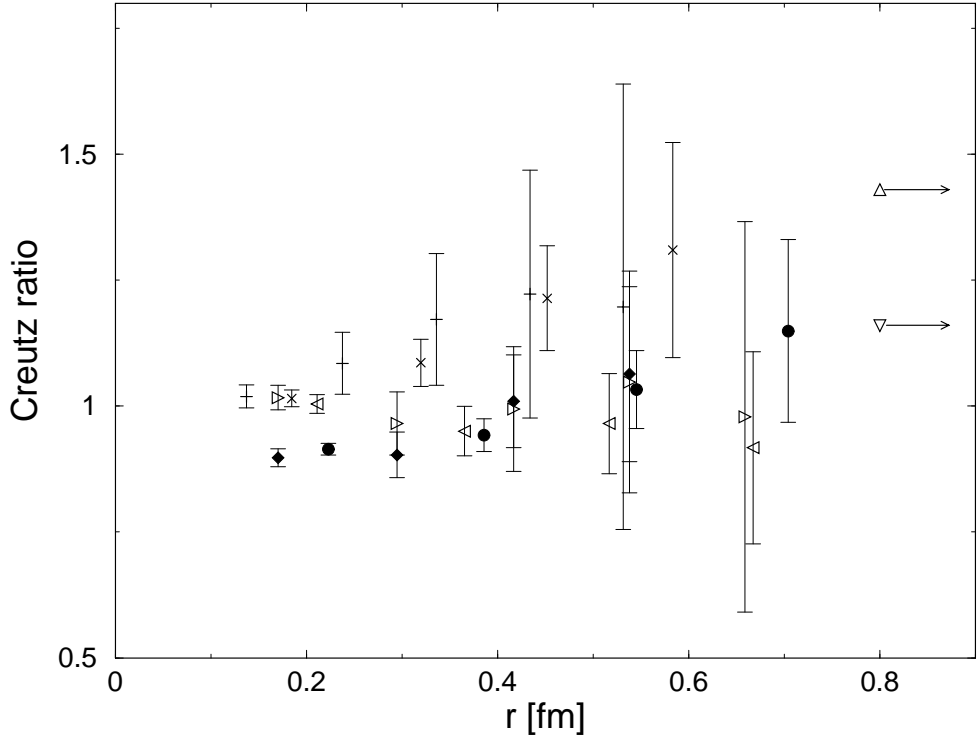}
}
\caption{Center-projected spatial Creutz ratios in units of the 
zero-temperature string tension, $\kappa (T\! \! =\! \! 0) a^2 $; 
$l\times l$ Creutz ratios are displayed at $r=\sqrt{l(l-1)} a(\beta )$ 
on the horizontal axis. Measurements were taken on a 
$12^3 \times N_t $ lattice. Shown are the temperatures $T=1.1T_C $ 
(open symbols; triangles pointing left correspond to $\beta =2.32, N_t =4$, 
whereas triangles pointing right correspond to $\beta =2.4, N_t =5$), 
$T=1.4T_C $ (filled symbols; diamonds correspond to $\beta =2.4, N_t =4$, 
whereas circles correspond to $\beta =2.3, N_t =3$), and $T=1.7T_C $ 
(crosses correspond to $\beta =2.48, N_t =4$, whereas 'x's correspond to 
$\beta =2.37, N_t =3$). For comparison, the spatial string tension 
extracted from full Wilson loops, as interpolated from data reported 
by Bali et al \cite{karsch}, is displayed: The triangle pointing downwards 
corresponds to $T=1.4T_C $, whereas the triangle pointing upwards 
corresponds to $T=1.7T_C $. The full spatial string tension at $T=1.1T_C $ 
is virtually indistinguishable from the zero-temperature value.}
\label{figa}
\end{figure}

The values obtained for the center-projected spatial Creutz ratios are 
summarized in Fig.~\ref{figa}, where they are compared with the 
high-precision data for the full spatial string tension of Bali et al 
\cite{karsch}. Since the temperatures used here and in \cite{karsch} do 
not coincide, an interpolation of the data points given in \cite{karsch} 
had to be carried out to arrive at the values depicted in Fig.~\ref{figa}.

Measurements were taken on a $12^3 \times N_t $ lattice, and
for each temperature, two values of the inverse coupling $\beta $ were
used. Note that there are two potential sources of scaling violations
in Fig.~\ref{figa}. On the one hand, center projection may destroy
the renormalization group scaling of the spatial string tension known
to occur when using the full configurations \cite{karsch}. This type 
of scaling violation would be a consequence, and thus a genuine indicator,
of vortex physics. On the other hand, the manner in which the data is
presented in Fig.~\ref{figa} also engenders additional scaling
violations to the extent in which Creutz ratios, which represent
difference quotients with increment $a(\beta )$, still deviate from
the derivatives they converge to as $a\rightarrow 0$. The authors have
elected to accept this slight disadvantage, since the presentation
of the data in Fig.~\ref{figa} is on the other hand well adapted to aid 
in the discussion below. Now, comparing the data obtained for different
$\beta $ at one temperature in Fig.~\ref{figa}, scaling violations
are evidently not significant as compared with the error bars. Namely,
values of Creutz ratios for two different choices of $\beta $ are
well described by a universal curve, better in fact than the error
bars would suggest. However, in view of the size of the error bars, 
which is due to the moderate statistics available to the authors, 
the data do not give very stringent evidence of correct renormalization 
group scaling; they are perhaps best described as being compatible 
with such scaling. 

Furthermore, the data seem to point towards a certain change in the 
dynamics generating the spatial string tension as the temperature is 
raised to values significantly above the deconfinement transition. 
At $T=1.1T_C $, the Creutz ratios are practically constant as a function 
of the Wilson loop size. This behavior of center-projected Wilson loops 
has been reported before in zero-temperature studies \cite{deb97} and has 
been dubbed ``precocious scaling''. Center projection truncates the 
short-range Coulomb behavior of full Wilson loops and one can read off the
asymptotic string tension already from $2\times 2$ Creutz ratios.

By contrast, this behavior does not seem quite as pronounced at temperatures
significantly above the deconfinement transition. Creutz ratios rise
as a function of loop size; it should however be mentioned that this
rise is much weaker than the usual Coulomb fall-off one obtains when 
using the full Yang-Mills configurations to evaluate the Creutz ratios.
Due to this variation with loop size, the asymptotic value of the full 
spatial string tension extracted from the data in \cite{karsch} is, in 
the case of $T=1.4T_C $, only reached by the Creutz ratio corresponding 
to the largest Wilson loops investigated; at $T=1.7T_C $, the asymptotic 
value is not quite reached even by the ratios derived from the most extended 
loops sampled, although it is within the error bars. While the error bars 
afflicting the Creutz ratios extracted from larger loops are sizeable, 
the rise as a function of loop size does seem to be significant, especially
as compared to the precocious scaling displayed at $T=1.1T_C$. Also
the difference between the values taken at $T=1.4T_C $ and $T=1.7T_C $
is compatible with the difference found for the full Wilson loops 
\cite{karsch}.

In view of their limited accuracy, the data depicted in Fig.~\ref{figa}
are perhaps best done justice by the statement that they do not allow
to negate the hypothesis of center dominance for the spatial string
tension in the deconfined phase. Certainly, no drastic deviation from
center dominance is apparent. However, more accurate studies of this 
question are clearly called for.

\section{Vortex percolation}
\subsection{Clustering of vortices}
As already mentioned above, there exists even in the deconfined phase
a substantial density of vortex intersection points on
the area spanned by two Polyakov loops \cite{tempv}. Thus, deconfinement
must be due more specifically to a correlation between these intersection
points, such that the distribution of points ceases to be sufficiently
random to generate an area law. As motivated in the introduction,
a correlation conducive to deconfinement would occur if vortices only
formed clusters smaller than some maximal size, i.e. if they ceased
to percolate. This would make the points appear in pairs separated by
less than the aforementioned maximal size, leading to a perimeter
law for the Polyakov loop correlator. In order to test whether this
type of mechanism is at work in connection with the Yang-Mills
deconfinement transition, it is necessary to measure the extension 
of vortex clusters.

Vortices constitute closed two-dimensional surfaces in four space-time
dimensions, or, equivalently, one-dimensional loops if one projects
down to three dimensions by taking a fixed time slice or a fixed
space slice of the (lattice) universe. Note that the term
space slice here is meant to denote the three-dimensional space-time 
one obtains by holding just one of the three space coordinates fixed.
Which particular coordinate is fixed is immaterial in view of spatial 
rotational invariance. In the following, specifically the extension of 
vortex line clusters in either time or space slices will be investigated. 
In this way, the relevant information is exhibited more clearly than by
considering the full two-dimensional vortex surfaces in four-dimensional
space-time.

Given a center-projected lattice configuration, the corresponding 
vortices can be constructed on the dual lattice in the fashion already
indicated in the introduction. As a definite example, consider a fixed time
slice. Then the vortices are described by lines made up of links on the 
dual lattice. Consider in particular a plaquette on the original lattice, 
lying e.g. in the $z=z_0 $ plane and extending from $x_0 $ to $x_0 +a$ 
and from $y_0 $ to $y_0 +a$, where $a$ denotes the lattice spacing. 
By definition, if the links making up this plaquette multiply to the 
center element $-1$, then a vortex pierces that plaquette. This means 
that a certain link on the dual lattice is part of a vortex; namely, the 
link connecting the dual lattice points $(x_0 +a/2 , y_0 +a/2 , z_0 -a/2 )$ 
and $(x_0 +a/2 , y_0 +a/2 , z_0 +a/2 )$.

Having constructed the vortex configuration on the dual lattice, one 
can proceed to define the vortex clusters. One begins by scanning the
dual lattice for a link which is part of a vortex. Starting from that
link, one tests which adjacent links, i.e. links which share a dual 
lattice site with the first link, are also part of the vortex. This
is repeated with all new members of the cluster until all links making 
up the cluster are found. In this way, it is possible to separate the 
different vortex clusters. 

\subsection{Extension of vortex clusters}
\label{secsp}
\begin{figure}[h]
\centerline{
\hspace{0.5cm}
\epsfysize=8cm
\epsffile{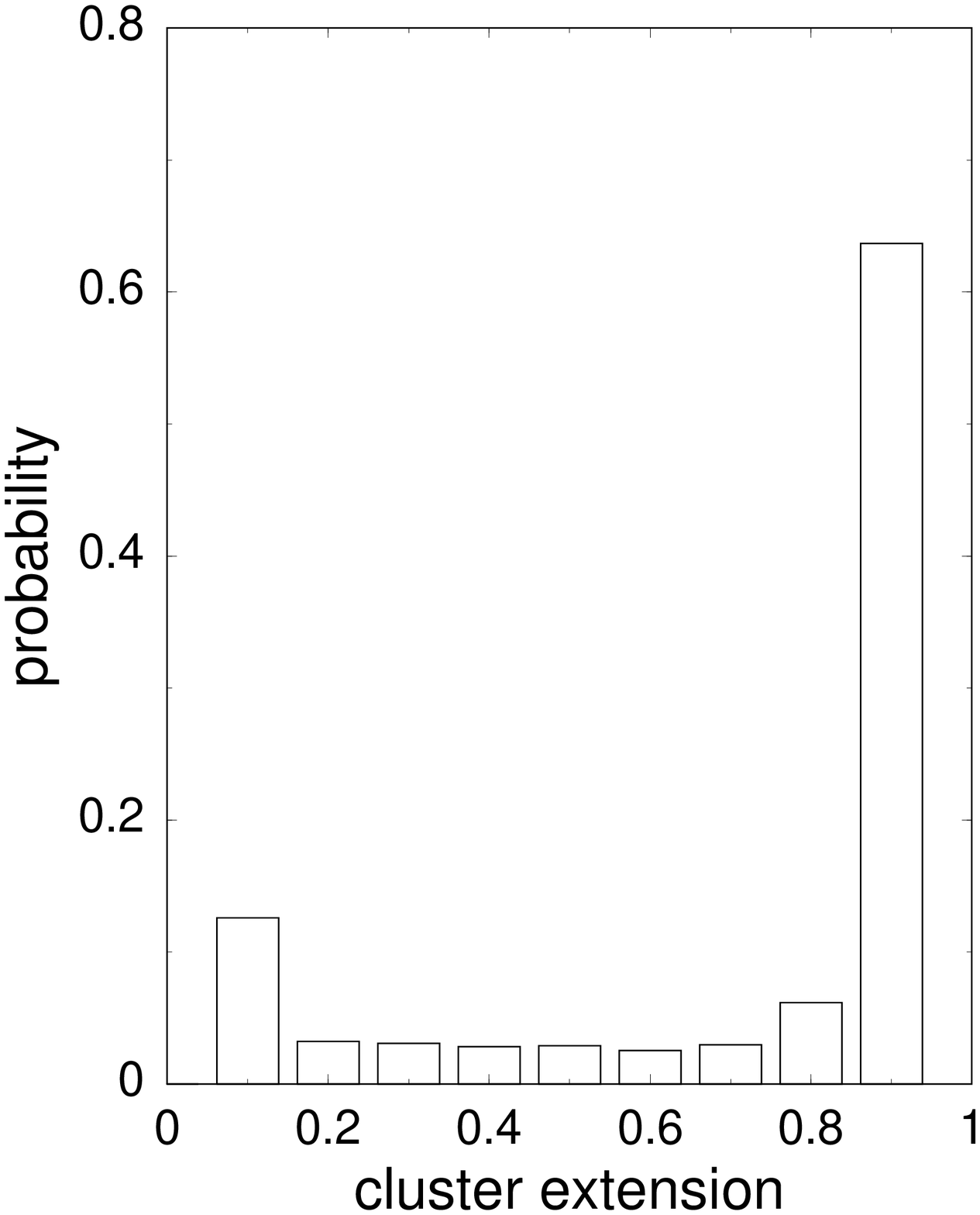}
\hspace{0.5cm}
\epsfysize=8cm
\epsffile{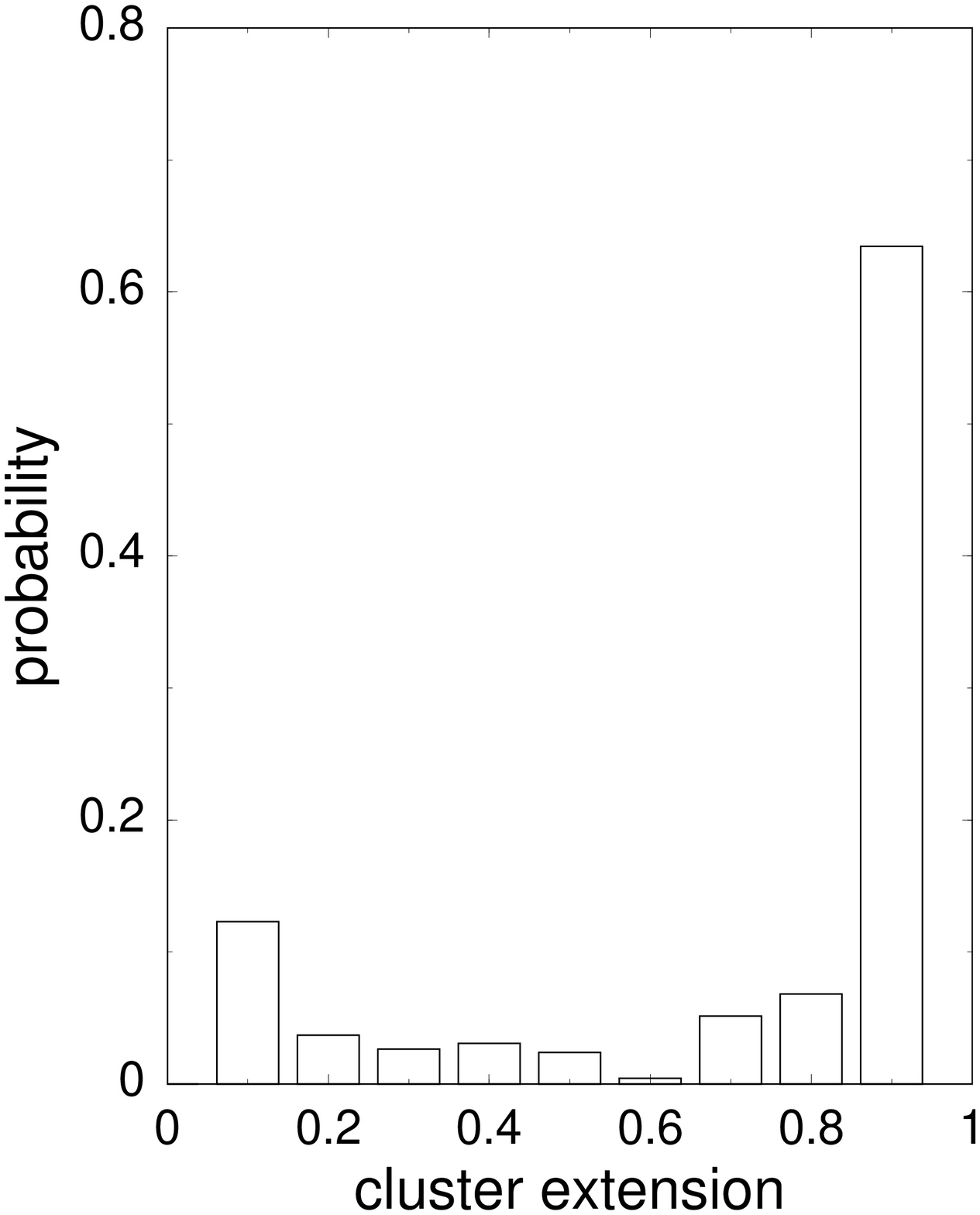}
}
\caption{Vortex material distributions in space slices of 
$12^3 \times N_t $ lattice universes
obtained as described in the text. Left: $12^3 \times 8$ lattice
at $\beta =2.4$, which is identified with $T=0.7T_C $. Right:
$12^3 \times 7$ lattice at $\beta =2.4$, which is identified with 
$T=0.8T_C $. The bins represent the percentage of vortex material 
organized into clusters of the corresponding extension. Extension 
is measured on the horizontal axis in units of the maximal extension
possible on a space slice of the given lattice, namely 
$\sqrt{2\cdot (12/2)^2 + (N_t /2)^2 } $ lattice spacings.}
\label{fig2}
\end{figure}

\begin{figure}[h]
\centerline{
\hspace{0.5cm}
\epsfysize=8cm
\epsffile{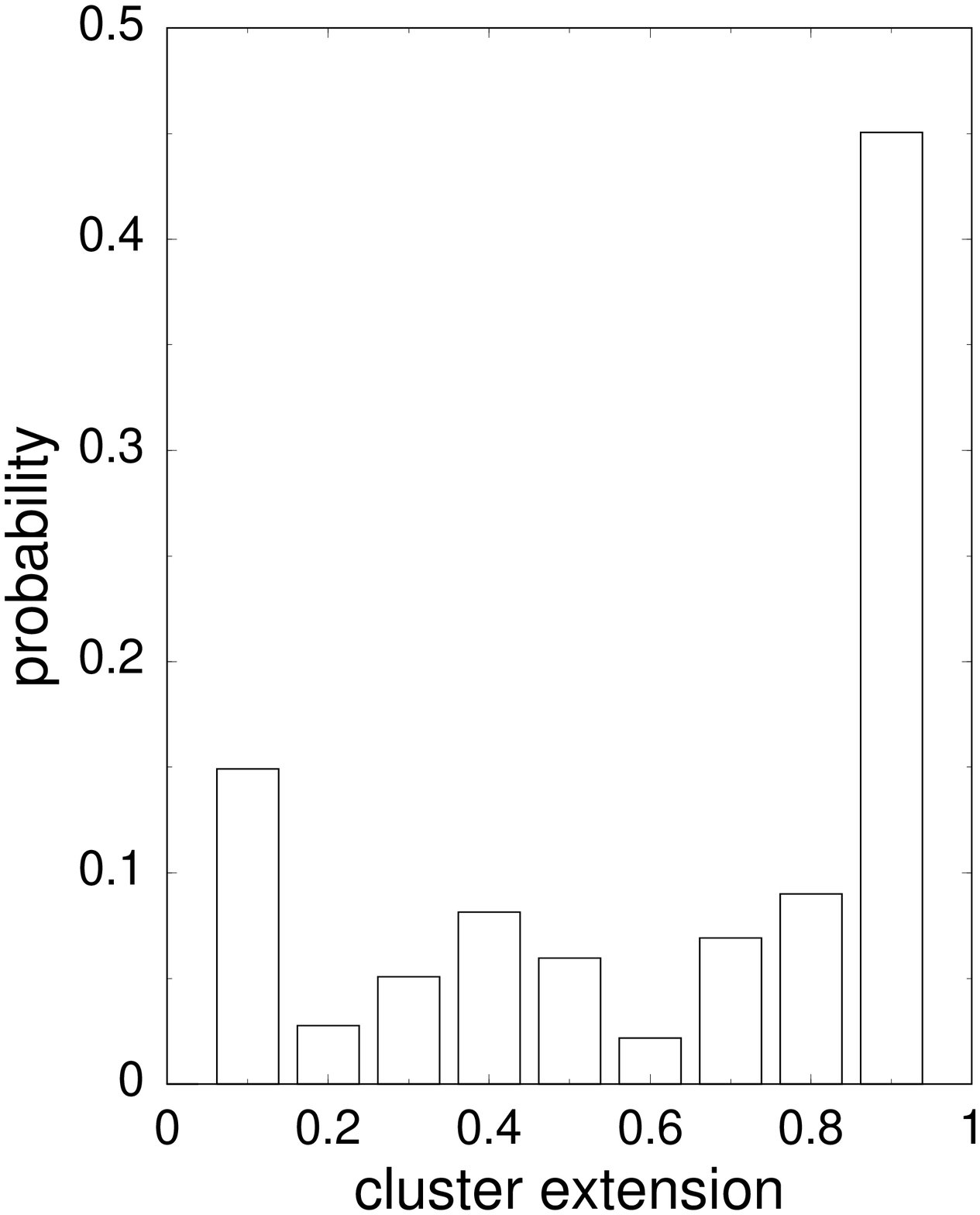}
\hspace{0.5cm}
\epsfysize=8cm
\epsffile{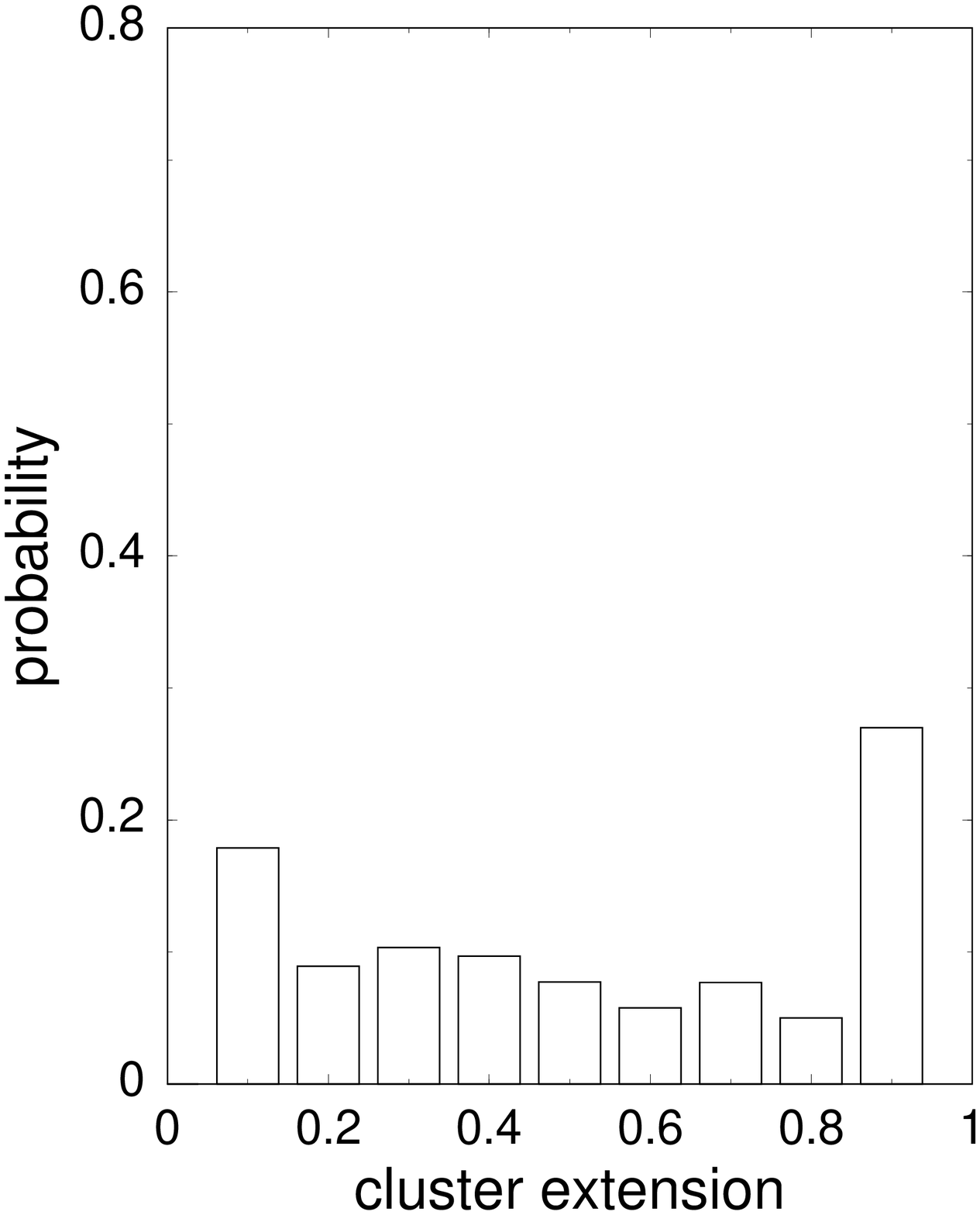}
}
\caption{Vortex material distributions as in Fig.~\ref{fig2}, at
different temperatures. Left: $12^3 \times 6$ lattice
at $\beta =2.4$, which is identified with $T=0.9T_C $. Right:
$12^3 \times 5$ lattice at $\beta =2.4$, which is identified with 
$T=1.1T_C $.}
\label{fig2a}
\end{figure}

\begin{figure}[h]
\centerline{
\hspace{0.5cm}
\epsfysize=8cm
\epsffile{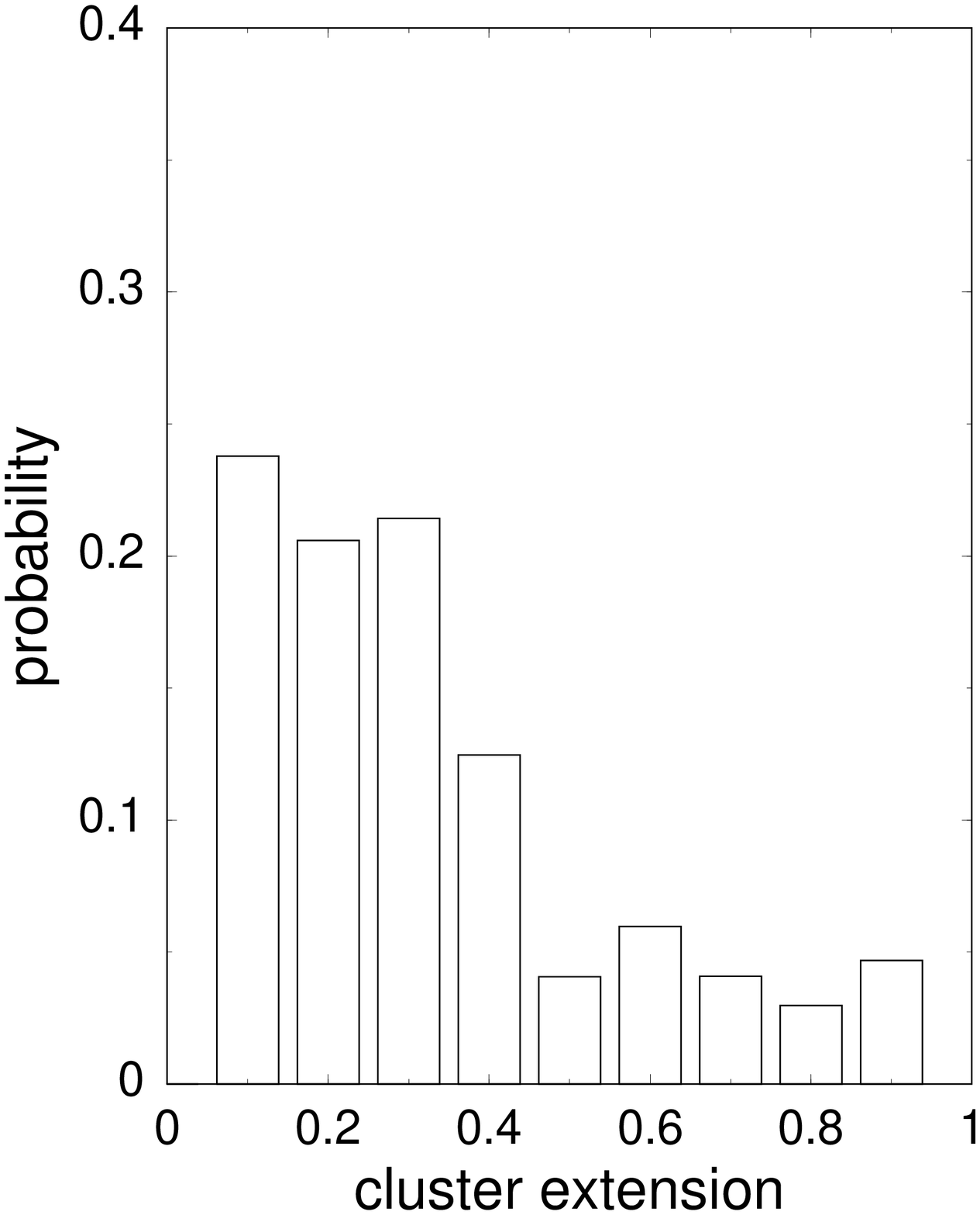}
\hspace{0.5cm}
\epsfysize=8cm
\epsffile{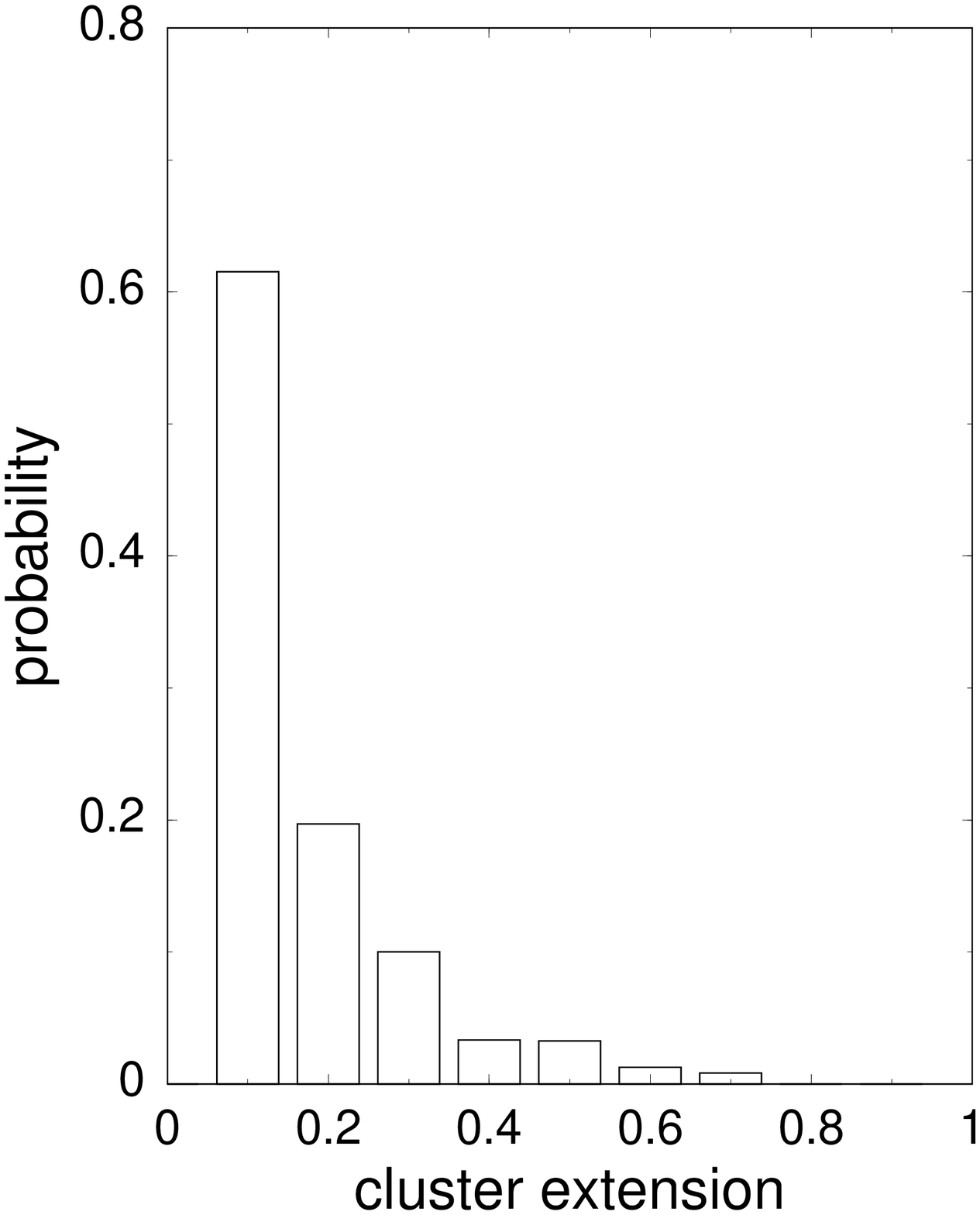}
}
\caption{Vortex material distributions as in Fig.~\ref{fig2}, at
different temperatures. Left: $12^3 \times 4$ lattice
at $\beta =2.4$, which is identified with $T=1.4T_C $. Right:
$12^3 \times 3$ lattice at $\beta =2.4$, which is identified with
$T=1.85T_C $.}
\label{fig3}
\end{figure}

\begin{figure}[h]
\centerline{
\epsfysize=9cm
\epsffile{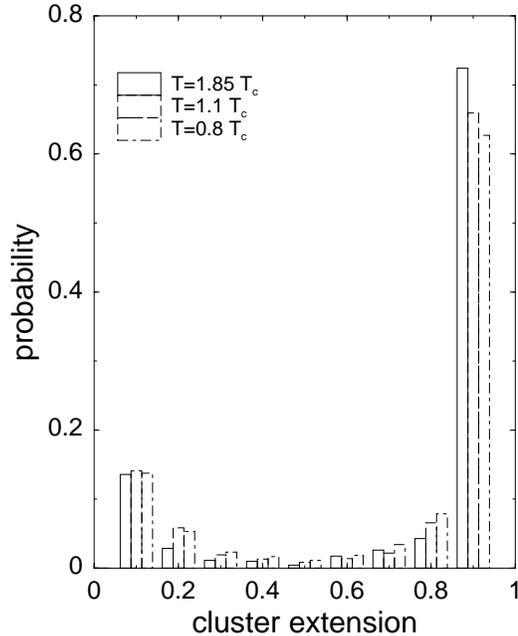}
}
\caption{Vortex material distributions analogous to Fig.~\ref{fig2}, but
taken from time slices of the $12^3 \times N_t $ lattice universe, again
at inverse coupling $\beta =2.4$. Bins corresponding to three different 
temperatures are shown simultaneously at each cluster extension; namely,
the case $N_t = 3$, which is identified with $T=1.85T_C $, the case
$N_t = 5$, which is identified with $T=1.1T_C $, and the case
$N_t = 7$, which is identified with $T=0.8T_C $.}
\label{fig4}
\end{figure}

Given the vortex clusters, their extensions
can be measured. Consider all pairs of links on a cluster and evaluate 
the space-time distance between each pair. The maximal such distance 
defines the extension of that cluster. In Figs.~\ref{fig2}-\ref{fig4}, 
histograms are displayed in which, for every cluster, the total number 
of links making up that cluster was added to the bin corresponding to 
the extension of the cluster.

The histograms were finally normalized such that the integral of the
distributions gives unity. Constructed in this way, the histograms
give a very transparent characterization of typical vortex 
configurations. The content of each bin represents the percentage of the 
total vortex length in the configurations, i.e. the available vortex 
material, which is organized into clusters of the corresponding extension. 
Accordingly, these distributions will be referred to as {\em vortex 
material distributions} in the following. In a percolating phase, the
vortex material distribution is peaked at the largest extension
possible on the lattice universe under consideration. Note that, due to
the periodic boundary conditions, this maximal extension e.g. on a 
$N_s \times N_s \times N_t $ space slice of the four-dimensional space-time
lattice is $\sqrt{(N_s /2)^2 + (N_s /2)^2 + (N_t /2)^2 } $ lattice spacings.
In a non-percolating phase, the vortex material distribution is
peaked at a finite extension independent of the size of the universe. 
Figs.~\ref{fig2}-\ref{fig3} pertain to space slices. Analogous results 
for time slices are summarized in Fig.~\ref{fig4}.

In space slices of the lattice universe, one observes a transition from a percolating to 
a non-percolating phase at the Yang-Mills deconfinement temperature.
Namely, in space slices, the vortex material distribution
is strongly peaked at the maximal possible extension as long as
the temperature remains below $T_C $; when the temperature
rises above $T_C $, however, the distribution becomes concentrated
at short lengths. The behavior near the deconfinement temperature
$T_C $ displayed in Figs.~\ref{fig2}-\ref{fig3} deserves more
detailed discussion. While the contents of the bin of maximal
extension fall sharply between $T=0.8T_C $ and $T=1.1T_C $, a residual 
one quarter of vortex material remains concentrated in loops of maximal
extension at the temperature identified as $T=1.1T_C $. This is too
large a proportion to let pass by without further consideration. The
authors have repeated the measurement at $T=1.1T_C $ on a larger, 
$16^3 \times 3$ lattice, and did not find a depletion of the bin of 
maximal extension. On the other hand, one should be aware that there is
a considerable uncertainty, of the order of $10\% $, in the overall
physical scale in these lattice experiments, affecting in particular
the identification of the deconfinement temperature $T_C $ itself.
These uncertainties were already mentioned in section \ref{centdo}
and are discussed in detail in \cite{tempv}. At the present level 
of accuracy, $T=1.1T_C $ cannot be considered significantly
separated from $T_C $; the authors cannot state with confidence that the
measurement formally identified with a temperature $T=1.1T_C $ must
unambiguously be associated with the deconfined phase.
Note that also in standard string tension measurements
via the Polyakov loop correlator, one does not attain a sharper signal
of the deconfinement transition if one uses comparable lattices and
statistics. Indeed, in \cite{tempv}, the authors still extracted a 
string tension of about $10\% $ of the zero-temperature value at the
temperature formally identified as $T=1.1T_C $.

In balance, the authors would argue that the percolation transition
in space slices does occur together with the deconfining transition, both
in view of the strong heuristic arguments connecting the two phenomena
in the vortex picture, and in view of the sharp change in the vortex
material distributions between $T=0.8T_C $ and $T=1.1T_C $. The latter
sharp change suggests that the vortex material distributions can in
practice be used as an alternative order parameter for the deconfinement
transition. When the vortices rearrange at the transition temperature to 
form a non-percolating phase, intersection points of vortices 
with planes containing Polyakov loop correlators occur in pairs less 
than a maximal distance apart. This leads to a perimeter law for the 
Polyakov loop correlator, implying deconfinement.

Consider now by contrast the vortex material distributions obtained
in time slices. According to Fig.~\ref{fig4}, these distributions
are strongly peaked at the maximal possible extension at all
temperatures, even above the deconfinement transition. Thus, vortex
line clusters in time slices always percolate; there is no marked
change in their properties as the temperature crosses $T_C $. 

Note that this entails no consequences for the behavior of the Polyakov 
loop correlator, since Polyakov loops do not lie within time slices. 
However, the persistence of vortex percolation into the deconfined phase 
when time slices are considered represents one way of understanding the 
persistence of a spatial string tension above $T_C $. Given percolation,
it seems plausible that intersection points of vortices with spatial 
Wilson loops continue to occur sufficiently randomly to generate an area 
law. There is another, complementary, way of understanding the spatial 
string tension which will be discussed in detail in the concluding section.

Note furthermore that Figs.~\ref{fig2}-\ref{fig4} taken together
imply that the vortices, regarded as two-dimensional surfaces in
four-dimensional space-time, percolate both in the confined and
the deconfined phases; this was also observed in \cite{bertl}.
Only by considering a space slice does one
filter out the percolation transition in the topology of the
vortex configurations. It should be emphasized that the percolation
of the two-dimensional vortex surfaces in four-dimensional space-time
in the deconfined phase does not negate the heuristic picture of
deconfinement put forward above. Given that vortex line clusters in 
space slices cease to percolate in the deconfined phase, intersection 
points of vortices with planes extending in one space and the time 
direction necessarily come in pairs less than a maximal distance apart, 
regardless of whether the different vortex line clusters do ultimately 
connect if one follows their world sheets into the additional spatial 
dimension. It is this pair correlation of the intersection points
which induces the deconfinement transition.

\subsection{Winding vortices in the deconfined phase}
\begin{figure}[h]
\centerline{
\epsfysize=9cm
\epsffile{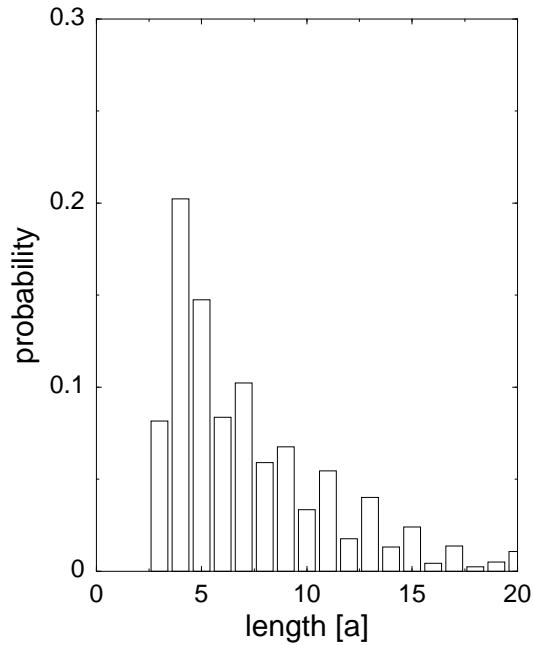}
}
\caption{Vortex material distributions in space slices of the lattice
universe as a function of total vortex line length contained in the
clusters. On the $12^3 \times N_t =3 $ lattice used, clusters with a 
length of an odd number of lattice spacings necessarily wind around the 
lattice in the Euclidean time direction. The inverse coupling in this
measurement was again set to the value $\beta =2.4$, implying that
this measurement is associated with a temperature of $T=1.85T_C $.
There is a residual, but insignificant, proportion of vortex clusters 
containing more than $20$ dual lattice links not displayed in the plot.}
\label{figb}
\end{figure}

\begin{figure}[h]
\centerline{
\epsfysize=7cm
\epsffile{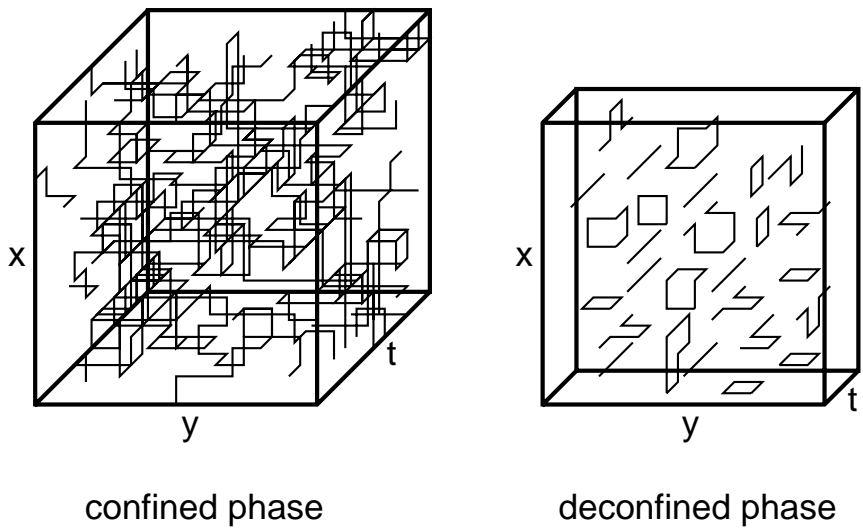}
\vspace{1cm}
}
\caption{Visualization of typical vortex configurations determining the
long-range physics of the confined and deconfined phases of Yang-Mills 
theory. Note that this is not a depiction of particular configurations 
found in lattice experiments; rather, it is the authors' interpretation 
of the measurements shown in Figs.~\ref{fig2}-\ref{figb} in terms of 
typical configurations dominating the Yang-Mills functional integral in 
the confined and deconfined phases. Shown are space slices of the lattice 
universe obtained by holding the $z$-coordinate fixed. Slicing the 
two-dimensional vortex surfaces present in four space-time dimensions 
yields one-dimensional loop configurations such as depicted.}
\label{phases}
\end{figure}

In order to gain a more detailed picture of the deconfined regime, it is
useful to carry out the following analysis. Consider again a space
slice of the lattice universe, in which vortex line clusters are short
in the deconfined regime. Consider in particular lattices of time
extension $N_t a$ with odd $N_t $, where $a$ is the lattice
spacing; in the following numerical experiment, $N_t =3$. On such a
lattice, measure vortex material distributions akin to the ones
described in the previous section, with one slight modification;
namely, define the bins of the histograms not by cluster extension,
but simply by the number of dual lattice links contained in the
clusters. It turns out that, in the deconfined phase, specifically at
$T=1.85T_C $, roughly $55\% $ of the vortex material is concentrated in 
clusters made up of an odd number of links, cf. Fig.~\ref{figb}. On a 
lattice with $N_t =3$, these are necessarily vortex loops which wind
around the lattice in (Euclidean) time direction by virtue of the
periodic boundary conditions, where the loops containing an odd number of
links larger than $3$ exhibit residual transverse fluctuations in the 
spatial directions, as also visualized in Fig.~\ref{phases} further below.

One thus obtains a quite specific characterization of the short vortices
appearing in the deconfined regime. This phase can evidently be visualized
largely in terms of short winding vortex loops with residual transverse
fluctuations if one considers a space slice of the lattice universe, 
cf. Fig.~\ref{phases}. Note that this picture also explains the partial 
vortex polarization observed in density measurements \cite{tempv}.

\section{Discussion and Outlook}
On the basis of the measurements shown in the preceding sections,
a detailed description of the confined and deconfined
phases of Yang-Mills theory in terms of center vortices emerges.
The typical vortex configurations present in the two phases are
visualized in Fig.~\ref{phases}. This picture allows an intuitive
understanding of the phenomenon of confinement as well as the
characteristics of the transition to the deconfined phase.
In the confined phase, vortex line clusters in space slices of the
lattice universe percolate. This allows intersection points of
vortices with planes containing Polyakov loop correlators to occur
sufficiently randomly to generate an area law. By contrast, in the
deconfined phase, typical vortex configurations in space slices of
the lattice universe are characterized by short vortex loops, to a
large part winding in the (Euclidean) time direction. This causes 
intersection points of vortices with planes containing Polyakov loop 
correlators to occur in pairs less than a maximal distance apart, leading 
to a perimeter law. Simple analytical model arguments clarifying the 
emergence of this qualitative difference were presented in the introduction.
The deconfinement phase transition in the vortex picture can thus
be understood as a transition from a percolating to a
non-percolating phase.

It should be emphasized that the percolation properties of vortices
focused on in the present work are more stringently related to
confinement than the polarization properties reported in \cite{tempv}.
There is a priori no direct logical connection between the observed
partial vortex polarization by itself and deconfinement. On the one hand,
even in presence of a significant polarization, confinement would persist 
as long as the vortex loops retain an arbitrarily large length, namely by 
winding sufficiently often around the (Euclidean) time direction before
closing. On the other hand, even in an ensemble with no polarization, 
deconfinement will occur if the vortices are organized into many small 
isolated clusters. Thus, vortex polarization should be viewed more as an
accompanying effect than the direct cause of deconfinement. Of course,
a correlation between the absence of percolation in space slices of
the lattice universe and vortex polarization is not surprising. If
fluctuations of vortex loops in the space directions are curtailed,
e.g. due to a phase containing many short vortices winding in the
time direction becoming favored (more about this below), then clearly
the connectivity of vortex clusters in the space direction is reduced
and they may cease to percolate. In this sense, polarization indirectly
can facilitate deconfinement. However, the percolation concept is related
much more directly and with much less ambiguity to the question of 
confinement. Ultimately, this is a consequence of a point already made 
in the introduction in connection with the heuristic models discussed 
there. Since the Wilson loop should be independent of the choice of 
area which one may regard it to span, it is conceptually sounder not 
to consider densities occuring on such areas, but the global topology 
of the vortices such as their linking number with the Wilson loop. 
The likelihood of a particular linking number occuring is strongly 
influenced by the connectivity of the vortex networks. Correspondingly, 
there is a clear signal of the phase transition in the vortex material 
distributions displayed in Figs.~\ref{fig2}-\ref{fig3}; these quantities 
can be used as alternative order parameters for the transition. By 
contrast, the vortex densities seem to behave smoothly across the 
deconfinement phase transition \cite{tempv}.

Turning to the spatial string tension, there are two complementary
ways to qualitatively account for its persistence in the deconfined
phase of Yang-Mills theory. One was already mentioned in 
section \ref{secsp}. If one considers a time slice of the lattice
universe, the associated vortex line configurations display no marked 
change of their clustering properties across the deconfinement 
transition. Even in the deconfined phase, vortex loops in time slices 
percolate. In view of this, it seems plausible that intersection points 
of vortices with spatial Wilson loops continue to occur sufficiently 
randomly to generate an area law. It should be noted, however, that this
percolation is qualitatively different from the one observed in the
confined phase in that it only occurs in the three space dimensions,
whereas the configurations are relatively weakly varying in the
Euclidean time direction. In other words, in the deconfined phase,
one finds a dimensionally reduced percolation phenomenon only visible
either in the full four space-time dimensions or in time slices thereof.

On the other hand, if one considers a space slice of the lattice universe, 
the deconfined phase is characterized to a large part by short vortex loops
winding in the time direction, cf. Fig.~\ref{phases}. However,
in this topological setup, such short vortices can pierce the area
spanned by a large spatial Wilson loop an odd number of times, even
far from its perimeter. This should be contrasted with the picture
one obtains for the Polyakov loop correlator. There, shortness of
vortices implies that their intersection points with the plane
containing the Polyakov loop correlator occur in pairs less than a
maximal distance apart. This leads to a perimeter law behavior of the
Polyakov loop correlator, i.e. deconfinement. For spatial Wilson
loops, this mechanism is inoperative due to the different topological
setting. On the contrary, in view of Fig.~\ref{phases}, if one
assumes the locations of the various winding vortices to be
uncorrelated, one obtains precisely the heuristic model of the
introduction, in which vortex intersection points are distributed
at random on the plane containing the spatial Wilson loop, leading to an
area law. Finite length vortex loops thus do not contradict the existence 
of a {\em spatial} string tension.

Of course, there is no reason to expect the locations of the winding 
vortices to be completely uncorrelated in the high-temperature Yang-Mills 
ensemble. In fact, comparing the values for the spatial string tension 
$\kappa_{s} $ from \cite{karsch} and the relevant density $\rho_{s} $
of vortex intersection points on planes extending in two spatial 
directions \cite{tempv}, the ratio $\kappa_{s} /\rho_{s} $ reaches values
$\kappa_{s} /\rho_{s} \approx 3 $ at $T\approx 2T_C $. 
This should be contrasted with the value
$\kappa =2\rho $ obtained in the model of random intersection points
discussed in the introduction. If one further takes into account that
a sizeable part of $\rho_{s} $ is still furnished by non-winding vortex
loops, cf. Fig.~\ref{figb}, then one should actually use the density
$\rho_{s}^{\prime } < \rho_{s} $ corresponding to winding vortices
only in the above consideration. This yields an even larger ratio
$\kappa_{s} /\rho_{s}^{\prime } $. Therefore, the winding vortices
in the deconfined phase seem to be subject to sizeable correlations.

Both of the above complementary mechanisms generating the spatial
string tension in the deconfined phase are qualitatively distinct
from the mechanism of confinement below $T_C $. In the space-slice
picture, this is obvious; a new class of configurations, namely
short vortex loops winding in the Euclidean time direction, induces
the spatial string tension. However, as already indicated further
above, also in the time-slice picture, the observed percolation is
qualitatively different from the one in the confined phase in that it
is dimensionally reduced. This qualitatively different origin of the
spatial string tension may provide a natural explanation for the
novel behavior detected in section \ref{centdo} for spatial Creutz ratios
at temperatures well inside the deconfined regime; namely, their rise
as a function of the size of the Wilson loops from which they are
extracted (as opposed to the precocious scaling observed at lower
temperatures). However, the detailed connection between the
abovementioned modified dynamics in the deconfined phase and the 
signal seen in the measurements of spatial Creutz ratios remains unclear.

While the relevant characteristics of the vortex configurations
in the different regimes were described in detail in this work, the 
present understanding of the underlying dynamics in the vortex picture 
is still tenuous. There are, however, indications that the deconfining
percolation transition can be understood in terms of simple
entropy considerations. Increasing the temperature implies
shortening the (Euclidean) time direction of the (lattice) universe.
This means that the number of possible percolating vortex
configurations decreases simply due to the reduction in space-time
volume\footnote{For example, a vortex surface extending into two space 
directions has a greatly reduced freedom of transverse fluctuation into 
the time direction. Note that if one thinks of such a fluctuating, 
fuzzy thin vortex surface in terms of a thick envelope, this amounts to 
stating that the thick vortex extending into two space directions simply 
does not fit into the space-time manifold anymore. To a certain extent,
the difference between these two (thin and thick vortex) pictures is
semantic. To state that a thick vortex does not fit into the space-time
manifold perpendicular to the time direction amounts to nothing but
the statement that the number of possible configurations of this
type has been reduced (to zero).}. At the same time as the number 
of possible percolating vortex clusters is reduced, the number of 
available short vortex configurations is enhanced by the emergence of a 
new class of short vortices at finite temperature, namely the vortices 
winding in time direction. In view of this, it seems plausible that a 
transition to a non-percolating phase is facilitated as temperature is 
raised.

There are two pieces of evidence supporting this explanation,
one of which was already given above. Namely, the deconfined phase
indeed contains a large proportion of short winding vortices, cf. 
Fig.~\ref{figb}. More than half of the vortex material is transferred
to the newly available class of short winding vortices in the deconfined
phase. The second piece of evidence 
is related to the behavior of stiff random surfaces in four
space-time dimensions; some of the authors plan to report on their
Monte Carlo investigation of these objects in an upcoming
publication. The model assumes that the vortices are random surfaces
associated with a certain action cost per unit area and a penalty for
curvature of the vortex surface. By construction, evaluating the partition
function of this model simply corresponds to counting the available vortex
configurations under certain constraints imposed by the action; namely,
the action cost per surface area effectively imposes a certain mean density 
of vortices, while the curvature penalty imposes an ultraviolet cutoff 
on the fluctuations of the vortex surfaces. Beyond this, no further 
dynamical information enters. It turns out that already this simple
model generates a percolation phase transition analogous to the one 
observed here for the center vortices of Yang-Mills theory. This
suggests that the deconfining percolation transition of center-projected 
Yang-Mills theory can be understood in similarly simple terms, without 
any need for detailed assumptions about the form of the full center vortex 
effective action.

\section{Acknowledgements}
Discussions with F.~Karsch and H.~Satz are gratefully acknowledged. 
K.L. also acknowledges the friendly hospitality of the members
of the KIAS, Korea, where a part of the numerical computations was 
carried out.

\end{document}